\documentclass[12pt]{article}

\usepackage{amsfonts, amsmath}

\newcommand{\be}{\begin{equation}} \newcommand{\ee}{\end{equation}}

\def\A{\ensuremath{\boldsymbol{A}}}

\begin{document}
\begin{center}
{\bf Quantum Theory at Planck Scale, Limiting Values, Deformed Gravity and Dark Energy Problem}\\
\vspace{5mm} A.E.Shalyt-Margolin \footnote{E-mail:
a.shalyt@mail.ru; alexm@hep.by}\\ \vspace{5mm} \textit{National
Center of Particles and High Energy Physics, Bogdanovich Str. 153,
Minsk 220040, Belarus}
\end{center}
\vspace{0.5cm}
\begin{abstract}
Within a theory of the existing fundamental length on the order of
Planck's a high-energy deformation of the General Relativity for
the space with horizon has been constructed. On this basis,
Markov's work of the early eighties of the last century has been
given a new interpretation to show that the heuristic model
considered by him may be placed on a fundamental footing. The
obtained results have been applied to solving of the dark energy
problem, making it possible to frame the following hypothesis: a
dynamic cosmological term is a measure of deviation from a
thermodynamic identity (the first law of thermodynamics) of the
high-energy (Planck's) deformation of Einstein equations for
horizon spaces in their thermodynamic interpretation.
\end{abstract}

\section{Introduction}
In the last decade numerous works devoted to a Quantum Field
Theory (QFT) at Planck's scale \cite{Planck1}--\cite{Planck3} have
been published(of course, the author has no pretensions of being
exhaustive in his references). This interest stems from the facts
that (i) at these scales it is expected to reveal the effects of a
Quantum Gravity (QG), and this still unresolved theory is
intriguing all the researchers engaged in  the field of
theoretical physics; (ii) modern accelerators, in particular LHC,
have the capacity of achieving the energies at which some QG
effects may be exhibited.
\\ Now it is clear that a Quantum Field Theory (QFT) at Planck's scales
undergoes changes associated with the appearance of additional
parameters related to  a minimal length (on the order of the
Planck's length). As this takes place, the corresponding
parameters are naturally considered as deformation parameters,
i.e. the related quantum theories are considered as a high-energy
deformation (at Planck's scales) of the well-known quantum field
theory.   The deformation is understood as an extension of a
particular theory by inclusion of one or several additional
parameters in such a way that the initial theory appears in the
limiting transition \cite{Fadd}.
\\ Of course, such a deformation must be adequately allowed for in
a gravitation theory. This work presents an approach to the construction
of a high-energy deformation of the General Relativity
in the case of horizon spaces, that will be termed here the Planck
deformation, and some inferences, in particular for the solution of
the dark energy problem.
\\ On the other hand, the dark material problem \cite{Sci1},
along with the dark energy problem \cite{Dar1}, is presently the
basic problem in modern fundamental physics, astrophysics, and
cosmology.  Whereas for a nature of the first real hypotheses have
been accepted already \cite{Hoop}, the dark energy still remains
enigmatic \cite{Quint1}--\cite{Phant1}. But it is the opinion of
most researchers that dark energy represents the energy of the
cosmic vacuum, its density being associated with the cosmological
term $\Lambda$ in Einstein's equation \cite{Gliner}--\cite{Wein1}.
In this respect an important reservation must be made – the point
is that most common is the term cosmological constant. Actually,
due to the Bianchi identities \cite{Wald}
 \be
 \nabla_\mu G^\mu{}_\nu = 0,
\label{bianchi}
 \ee
where $G^\mu{}_\nu$ -- Einstein equations
 \be G^\mu{}_\nu \equiv
R^\mu{}_\nu - \frac{1}{2} \, \delta^\mu{}_\nu R = \frac{8 \pi
G}{c^4} \, \left[T^\mu{}_\nu + \frac{c^4 \Lambda}{8 \pi G} \,
\delta^\mu{}_\nu \right] \label{einstein} \ee
the energy-momentum tensor $T^\mu{}_\nu $ (energy-momentum density tensor)
remains covariantly valid
\be \nabla_\mu T^\mu{}_\nu = 0.
\label{colaw0}
 \ee
From whence it directly follows that the cosmological term $\Lambda$ is a constant.
\\ But, as has been rightly noted in several publications
 (e.g., \cite{varyingL}--\cite {Var-CC2}), conservation laws
(\ref{colaw0}) are only regulating the energy-momentum exchange
between the  field sources and gravitational field and are liable
to be violated if an independent energy source is existent in the
Universe. Such a source may be associated with a time-varying
cosmological term. So, in this case it is reasonable to consider
$\Lambda=\Lambda(t)$.
\\ Then the Bianchi identity (\ref{bianchi})
is replaced by the "generalized Bianchi identity"  \cite{Var-CC1}
 \be \nabla_\mu \left[T^\mu{}_\nu
+ \Lambda^\mu{}_\nu \right] = 0, \label{colaw1} \ee
 where
$\Lambda^\mu{}_\nu = \varepsilon_\Lambda \delta^\mu{}_\nu$
is some energy-momentum tensor (referred to as the dark energy-momentum
tensor \cite {Var-CC1}) related to the cosmological term,
where \be \varepsilon_\Lambda = \frac{c^4 \Lambda}{8 \pi G}
\label{deden} \ee is the corresponding energy density.
\\The present work is a natural continuation of the previous papers of the author.
The idea that a quantum theory at the Planck scales must involve
the fundamental length has been put forward in the works devoted
to a string theory fairly a long time ago \cite{Ven1}. But since
it is still considered to be a tentative theory, some other
indications have been required. Fortunately, by the present time
numerous publications have suggested the appearance of the
fundamental length in the Early Universe with the use of various
approaches \cite{GUPg1}--\cite{Ahl1}. Of particular importance is
the work \cite{GUPg1}, where on the basis of a simple gedanken
experiment it is demonstrated that, with regard to the
gravitational interactions (Planck's scales) exhibited in the
Early Universe only, the Heisenberg Uncertainty Principle should
be extended to the Generalized Uncertainty Principle
\cite{Ven1}--\cite{Ahl1}that in turn is bound to bring forth the
fundamental length on the order of Planck's length. The advent of
novel theories in physics of the Early Universe is associated with
the introduction of new parameters, i.e. with a deformation of the
well-known theories. Of course, in this case Heisenberg Algebra is
subjected to the corresponding deformation too. Such a deformation
may be based on the Generalized Uncertainty Principle (GUP)
\cite{Magg}--\cite{Kempf} as well as on the density matrix
deformation \cite{shalyt1}--\cite{shalyt13}.
\\At the same time, the above-mentioned new deformation parameters
so far have not appeared in gravity despite the idea that they
should. The situation is that no evident efforts have been
undertaken to develop the high-energy (Planck's scale) gravity
deformations including the deformation parameters introduced in a
Quantum Theory of the Early Universe.
\\ In this paper, with   GUP held true, the possibility
for the high-energy gravity deformation is considered for a
specific case of Einstein's equations. As this takes place, the
parameter $\alpha$ appearing in the Quantum Field Theory (QFT)
with the UV cutoff (fundamental length) produced by the density
matrix deformation is used. There is no discrepancy of any kind as
the deformation parameter in the GUP-produced Heisenberg algebra
deformation is quite naturally expressed in terms of $\alpha$, and
this will be shown later (Section 2).  Besides, by its nature,
$\alpha$ is better applicable to study the high-energy deformation
of General Relativity because it is small, dimensionless (making
series expansion more natural), and the corresponding
representation of the  Einstein's equations in its terms or its
deformation appear simple.
 Structurally, the paper is as follows.
In Section 2 the approaches to the deformation of a quantum theory
at the Planck scales are briefly reviewed. In Section 3 it is
demonstrated that an heuristic approach to the high-energy
deformation of the General Relativity, considered earlier in
\cite{Mark1}, may be understood within a theory involving the
fundamental length. In Section 4 it is shown how to interpret
quantum corrections for the thermodynamic characteristics of black
holes considering the deformation of a theory with the fundamental
length. Essentially new results are presented in Sections 5 and 6.
A thermodynamic description of the General Relativity is used. The
possibility for the high energy deformation of Einstein's
equations is discussed within the scope of both equilibrium
thermodynamics and non-equilibrium thermodynamics. In the latter
case the approach is contemplated only in terms of a nature of the
cosmological term. Moreover, in this case a more precise
definition for the dependence of this term on the deformation
parameter is possible. In Section 7 the derived results are
applied to solve the dark energy problem.  Based on the results
obtained, in Conclusion the following hypothesis is framed:
\\{\bf a dynamic cosmological term is a measure of deviation
from the thermodynamic identity (the first law of thermodynamics)
of the high-energy (Planck's) deformation of Einstein equations
for horizon spaces in their thermodynamic interpretation}.
\section{Quantum Theory at Planck's Scale}
In the last twenty years the researchers have come to the
understanding that studies of the Early Universe physics
(extremely high – Planck's energies) necessitate changes in the
fundamental physical theories, specifically quantum mechanics and
quantum field theory. Inevitably a fundamental length should be
involved in these theories \cite{Gar1}--\cite{Magg1}. This idea
has been first suggested by a string theory \cite{Ven1}. But it is
still considered to be a tentative theory without the experimental
status and merely an attractive model. However, the fundamental
length has been involved subsequently in more simple and natural
considerations \cite{GUPg1}.
\\The main approach to framing of Quantum Mechanics with fundamental
length (QMFL) and Quantum Field Theory with fundamental length
(QFTFL) (or with Ultraviolet (UV) cutoff) is that associated with
the Generalized Uncertainty Principle (GUP)
\cite{Ven1}--\cite{Kempf}:
\begin{equation}\label{GUP1}
\triangle x\geq\frac{\hbar}{\triangle p}+\alpha^{\prime}
l_{p}^2\frac{\triangle p}{\hbar}.
\end{equation}
 with the corresponding Heisenberg
algebra deformation produced by this principle
\cite{Magg}--\cite{Kempf}. \\Besides, in the works by the author
\cite{shalyt1}--\cite{shalyt10} an approach to the construction of
QMFL has been developed with the help of the deformed density
matrix, the density matrix deformation in QMFL being a starting
object called the density pro-matrix $\rho(\alpha)$ and
deformation parameter (additional parameter)
\begin{equation}\label{D1}
\alpha=l_{min}^{2}/x^{2},
\end{equation}
where $x$ is the measuring scale, $l_{min}\sim l_{p}$ and
$0<\alpha\leq 1/4$ \cite{shalyt1},\cite{shalyt2}.
\\
\\The explicit form of the above-mentioned deformation gives the
exponential ansatz
\begin{equation}\label{U26S}
\rho^{*}(\alpha)=exp(-\alpha)\sum_{i}\omega_{i}|i><i|,
\end{equation}
where all $\omega_{i}>0$ are independent of $\alpha$ and their sum
is equal to 1.
\\ In the corresponding deformed Quantum Theory (denoted as $QFT^{\alpha}$)
for average values we have
\begin{equation}\label{U26cS}
<B>_{\alpha}=exp(-\alpha)<B>,
\end{equation}
where $<B>$ - average in well-known QFT
\cite{shalyt6},\cite{shalyt7}.
 All the variables associated with the considered $\alpha$ -
deformed quantum field theory  are hereinafter marked with the
upper index $^{\alpha}$.
\\ Note that the deformation parameter
$\alpha$ is absolutely naturally represented as a ratio between
the squared UV and IR limits
\begin{equation}\label{U26dS}
\alpha=(\frac{UV}{IR})^{2},
\end{equation}
where UV is fixed and IR is varying.
\\It should be noted \cite{shalyt13} that in a series of the author's
works \cite{shalyt1}--\cite{shalyt10} a minimal
$\alpha$-deformation of QFT has been formed. By "minimal" it is
meant that no space-time noncommutativity was required, i.e. there
was no requirement for noncommutative operators associated with
different spatial coordinates
\begin{equation}\label{Concl1}
[X_{i},X_{j}]\neq 0, i\neq j.
\end{equation}
However, all the well-known deformations of QFT associated with
GUP (for example, \cite{Magg}--\cite{Kempf}) contain
(\ref{Concl1}) as an element of the corresponding deformed
Heisenberg algebra. Because of this, it is necessary to extend (or
modify) the above-mentioned minimal $\alpha$-deformation of QFT
--$QFT^{\alpha}$ \cite{shalyt1}--\cite{shalyt10} to some new
deformation $\widetilde{QFT}^{\alpha}$ compatible with GUP, as it
has been noted in \cite{shalyt13}. We can easily show that  QFT
parameter of deformations associated with GUP may be expressed in
terms of the parameter $\alpha$ that has been introduced in the
approach associated with the density matrix deformation
\cite{shalyt-aip},\cite{shalyt-entropy2}. Here the notation of
\cite {Kim2} is used. Then
\begin{equation} \label{comm1}
[\vec{x}, \vec{p}]=i\hbar(1+\beta^2\vec{p}^2+...)
\end{equation}
and
\begin{equation}\label{comm2}
\Delta x_{\rm min}\approx
\hbar\sqrt{\beta}\sim l_{p}.
\end{equation}
Then from (\ref{comm1}),(\ref{comm2}) it follows that $\beta\sim
{\bf 1/p^{2}}$,  and for $x_{\rm min}\sim l_{p}$, $\beta$
corresponding to $x_{\rm min}$ is nothing else but
\begin{equation}\label{comm3}
\beta\sim  1/P_{pl}^{2},
\end{equation}
where $P_{pl}$ is Planck's momentum: $P_{pl}= \hbar/l_{p}$.
\\In this way $\beta$ is changing over the following interval:
\begin{equation}\label{comm4}
\lambda/P_{pl}^{2}\leq \beta<\infty,
\end{equation}
where $\lambda$  is  a numerical factor  and the second member in
(\ref{comm1}) is accurately reproduced in momentum representation
(up to the numerical factor) by $\alpha=l^{2}_{min}/l^{2}\sim
l^{2}_{p}/l^{2}=p^{2}/P_{pl}^{2}$
\begin{equation} \label{comm5}
[\vec{x},\vec{p}]=i\hbar(1+\beta^2\vec{p}^2+...)=i\hbar(1+a_{1}\alpha+a_{2}\alpha^{2}+...).
\end{equation}

\section{Density Limit, Fundamental Length, and Deformed Theories}
It should be noted that deformations at Planck's scales (the early
Universe) have been considered implicitly long before the works
\cite{Ven1}--\cite{Ahl1},\cite{Magg}--\cite{Kempf},\cite{shalyt1}--\cite{shalyt10}
devoted to quantum mechanics with the fundamental length. Let us
dwell on the work \cite{Mark1}, where it is assumed that "by the
universal decree of nature a quantity of the material density
$\varrho$ is always bounded by its upper value given by the
expression that is composed of fundamental constants"
(\cite{Mark1}, p.214):
\begin{equation}\label{Mark1}
\varrho\leq\varrho_{p}=\frac{c^{5}}{G^{2}\hbar},
\end{equation}
with $\varrho_{p}$ as "Planck's density".
\\ It is clearly seen that, proceeding from the involvement
of the fundamental length on the order of the Planck's
$l_{min}\sim l_{p}$, one can obtain $\varrho_{p}$  (\ref{Mark1})
up to a constant. Indeed, within the scope of GUP (\ref{GUP1})
(but not necessarily) we have
 $l_{min}=2\alpha^{\prime}l_{p}$  and then,
 as it has been shown in \cite{shalyt3},
(\ref{GUP1}) may be generalized to the corresponding relation of
the pair "energy - time" as follows:
\begin{equation}\label{Mark2}
\Delta t\geq\frac{\hbar}{\triangle
E}+\alpha^{\prime}t_{p}^{2}\frac{\triangle E}{\hbar}.
\end{equation}
This directly suggests the existence of the "minimal time"
$t_{min}=2\alpha^{\prime}t_{p}$ and of the "maximal energy"
corresponding to this minimal time $E_{max}\sim E_{p}$ .
\\Clearly, this maximal energy is associated with some "maximal
mass"
$M_{max}$:
\begin{equation}\label{Mark3}
E_{max}=M_{max}c^{2}, M_{max}\sim M_{p}
\end{equation}
Whence, considering that the existence of a minimal
three-dimensional volume $V_{min}=l^{3}_{min}\sim V_{p}=l^{3}_{p}$
naturally follows from the existence  of $l_{min}\sim l_{p}$, we
immediately arrive at the "maximal density"  $\varrho_{p}$
(\ref{Mark1}) but only within the factor determined by
$\alpha^{\prime}$
\begin{equation}\label{Mark4}
\frac{M_{max}}{V_{min}}=\varrho_{max}\sim \varrho_{p}.
\end{equation}
Actually, the quantity
\begin{equation}\label{Mark4.1}
\wp_{\varrho}=\varrho/\varrho_{p}\leq 1
\end{equation}
in \cite{Mark1} is the deformation parameter as it is used to
construct the deformation of Einstein's equation
(\cite{Mark1},formula (2)):
\begin{equation}\label{Mark5}
R^{\nu}_{\mu}-\frac{1}{2}R\delta^{\nu}_{\mu}=\frac{8\pi
G}{c^{4}}T^{\nu}_{\mu}(1-\wp_{\varrho}^{2})^{n}-\Lambda\wp_{\varrho}^{2n}\delta^{\nu}_{\mu},
\end{equation}
where  $n\geq 1/2$, $T^{\nu}_{\mu}$--energy-momentum tensor, $\Lambda$-- cosmological  constant.
The case of the parameter $\wp_{\varrho}\ll 1$ or $\varrho\ll \varrho_{p}$
correlates with the classical Einstein equation, and the case when
$\wp_{\varrho}= 1$ --  with the de Sitter Universe. In this way
(\ref{Mark5}) may be considered as $\wp_{\varrho}$-deformation
of  the General Relativity.
\\ As it has been noted before, the existence of a maximal
density directly, up to a constant, follows from the existence of
a fundamental length (\ref{Mark1}). It is clear that the
corresponding deformation parameter $\wp_{\varrho}$
(\ref{Mark4.1}) may be obtained from the deformation parameter
$\alpha$ (\ref{D1}). In fact, since $\alpha=l_{min}^{2}/x^{2}$, we
have
\begin{equation}\label{Mark6}
\alpha^{3/2}=\frac{l_{min}^{3}}{x^{3}}\sim \frac{V_{min}}{V},
\end{equation}
where $V$ is the three-dimensional volume associated with  the linear dimension $x$.
As in the "energy" representation
\begin{equation}\label{Mark7}
\alpha^{1/2}=E/E_{max}
\end{equation}
and considering that   $E_{max}\sim E_{p}$, and $V_{min}=l^{3}_{min}\sim
V_{p}=l^{3}_{p}$, we get
\begin{equation}\label{Mark8}
\wp_{\varrho}\sim
(E/E_{max})(V_{min}/V)=\frac{E/V}{E_{max}/V_{min}}=\frac{\varrho}{\varrho_{max}}=\alpha^{2}.
\end{equation}
Of course, the proportionality factor in (\ref{Mark8}) is model dependent.
 Specifically, if QMFL is related to GUP, this factor is depending on
 $\alpha^{\prime}$ (\ref{GUP1}).
 But the deformation parameters $\wp_{\varrho}$ and
 $\alpha$ are differing considerably: the limiting value $\wp_{\varrho}=1$
 is obviously associated with singularity, whereas originally
 by the approach involving the density matrix deformation
 \cite{shalyt2}--\cite{shalyt4},\cite{shalyt9},\cite{shalyt10}
no consideration has been given to the deformation parameter
 $\alpha=1$ associated with singularity, as it is ignored
 in accordance with the main definition of the
 $\alpha$-deformed density matrix
 \begin{equation}\label{Mark9}
Sp[\rho(\alpha)]\approx\frac{1}{2}+\sqrt{\frac{1}{4}-\alpha}.
\end{equation}
Because the parameter $\alpha$, as distinct from $\wp_{\varrho}$, is small
(and $\alpha^{2}$  is corresponding to $\wp_{\varrho}$), a series expansion by it is possible.
\\ So, $\wp_{\varrho}$-deformation of  the  General
Relativity may be interpreted as $\alpha$-deformation.
\\ In what follows it is demonstrated that the results presented
in this Section may be extended quite unexpectedly: in a
sufficiently general case we can treat a high-energy (Planck)
$\alpha$-deformation of the  General Relativity and hence an
heuristic model (\ref{Mark5})(formula (2)\cite{Mark1}) for
$\wp_{\varrho}$-deformation of  the  General Relativity may be
placed on a more fundamental footing.

\section{Gravitational Thermodynamics
in Low and High Energy and Deformed Quantum Theory}
In the last
decade a number of very interesting works have been published. We
can primary name the works \cite{Padm1}--\cite{Padm12}, where
gravitation, at least for the â spaces with horizon, is directly
associated with thermodynamics and the results obtained
demonstrate a holographic character of gravitation. Of the
greatest significance is a pioneer work  \cite{Jac1}. For black
holes the association has been first revealed in
\cite{Bek1},\cite{Hawk3}, where related the black-hole event
horizon temperature to the surface gravitation. In \cite{Padm11},
has shown that this relation is not accidental and may be
generalized for the spaces with horizon.  As all the foregoing
results have been obtained in a semiclassical approximation, i.e.
for sufficiently low energies, the problem arises: how these
results are modified when going to higher energies. In the context
of this paper, the problem may be stated as follows: since we have
some infra-red (IR) cutoff $l_{max}$ and ultraviolet (UV) cutoff
$l_{min}$, we naturally have a problem how the above-mentioned
results on Gravitational Thermodynamics are changed for
\begin{equation}\label{GT1}
l \rightarrow l_{min}.
\end{equation}
Sections 2 and 3 of this paper show that the results are dependent
on the deformation parameter $\alpha$ (\ref{D1}),(\ref{U26dS})
that in the accepted notation is of the form
\begin{equation}\label{GT2}
\alpha=\frac{l_{min}^{2}}{l^{2}}.
\end{equation}
In fact, in several papers \cite{acs}--\cite{Nou} it has been
demonstrated that thermodynamics and statistical mechanics of
black holes in the presence of GUP (i.e. at high energies) should
be modified.  To illustrate, in \cite{Park} the Hawking
temperature modification has been computed in the asymptotically
flat space in this case in particular. It is easily seen that in
this case the deformation parameter $\alpha$ arises naturally.
Indeed, modification of the Hawking temperature is of the
following form \cite{acs},\cite{Cava:04}\cite{Park}:
\begin{equation}\label{GT3}
T_{GUP}=(\frac{d-3}{4\pi})\frac{\hbar r_{+}}{2\alpha^{\prime
2}l^{2}_{p}}[1-(1-\frac{4\alpha^{\prime 2}l_{p}^{2}}{
r_{+}^{2}})^{1/2}],
\end{equation}
where $d$ is the space-time dimension, and $r_+$ is the
uncertainty in the emitted particle position by the Hawking
effect, expressed as
\begin{equation}\label{GT4}
\Delta x_i \approx r_+
\end{equation}
and being nothing else but a radius of the event horizon;
$\alpha^{\prime}$ -- dimensionless constant from GUP. But as we
have $2\alpha^{\prime}l_{p}=l_{min}$, in terms of $\alpha$
 (\ref{GT3}) may be written in a natural way as follows:
\begin{equation}\label{GT5}
T_{GUP}=(\frac{d-3}{4\pi})\frac{\hbar \alpha^{-1}_{r_{+}}
}{\alpha^{\prime}l_{p}}[1-(1-\alpha_{r_{+}})^{1/2}],
\end{equation}
where $\alpha_{r_{+}}$- parameter $\alpha$ associated with the
IR-cutoff $r_{+}$. In such a manner $T_{GUP}$ is only dependent on
the constants including the fundamental ones and on the
deformation parameter $\alpha$.
\\The dependence of the black hole entropy on $\alpha$ may be derived
in a similar way. For a semiclassical approximation of the
Bekenstein-Hawking formula \cite{Bek1},\cite{Hawk3}
\begin{equation}\label{GT6}
S=\frac{1}{4}\frac{A}{l^{2}_{p}},
\end{equation}
where $A$ -- surface area of the event horizon, provided the
horizon event has radius $r_+$, then $A\sim r^{2}_+$ and
(\ref{GT6}) is clearly of the form
\begin{equation}\label{GT6.1}
S=\sigma \alpha^{-1}_{r_{+}},
\end{equation}
where $\sigma$ is some dimensionless denumerable factor. The
general formula for quantum corrections \cite{mv} given as
\begin{equation}\label{GT6.2}
S_{GUP} =\frac{A}{4l_{p}^{2}}-{\pi\alpha^{\prime 2}\over 4}\ln
\left(\frac{A}{4l_{p}^{2}}\right) +\sum_{n=1}^{\infty}c_{n}
\left({A\over 4 l_p^2} \right)^{-n}+ \rm{const}\;,
\end{equation}
where the expansion coefficients $c_n\propto \alpha^{\prime
2(n+1)}$ can always be computed to any desired order of accuracy
\cite{mv}, may be also written as a power series in
$\alpha^{-1}_{r_{+}}$   (or  Laurent series in $\alpha_{r_{+}}$)
\begin{equation}\label{GT6.3}
S_{GUP}=\sigma \alpha^{-1}_{r_{+}}-{\pi\alpha^{\prime 2}\over
4}\ln (\sigma \alpha^{-1}_{r_{+}}) +\sum_{n=1}^{\infty}(c_{n}
\sigma^{-n}) \alpha^{n}_{r_{+}}+ \rm{const}
\end{equation}
Note that here no consideration is given to the restrictions on
the IR-cutoff
\begin{equation}\label{GT7}
l\leq l_{max}
\end{equation}
and to those corresponding the extended uncertainty principle
(EUP)\cite{Park} or symmetric generalized uncertainty principle
(SGUP)\cite{Kim1} that leads to a minimal momentum.
\\A black hole is a specific example of the space with horizon.
It is clear that for other horizon spaces \cite{Padm11} a similar
relationship between their thermodynamics and  the deformation
parameter $\alpha$ should be exhibited.
\\Quite recently,
in a series of papers, and specifically in
\cite{Padm3}--\cite{Padm9}, it has been shown that Einstein
equations may be derived from the surface term of the GR
Lagrangian, in fact containing the same information as the bulk
term.
\\It should be noted that Einstein's
equations [at least for space with horizon] may be obtained from
the proportionality of the entropy and horizon area together with
the fundamental thermodynamic relation connecting heat, entropy,
and temperature \cite{Jac1}. In fact \cite{Padm3}-- \cite{Padm10},
this approach has been extended and complemented by the
demonstration of holographicity  for the gravitational action (see
also \cite{Padm11}).And in the case of Einstein-Hilbert gravity,
it is possible to interpret Einstein's equations as the
thermodynamic identity \cite{Padm12}:
\begin{equation}\label{GT8}
TdS = dE + PdV.
\end{equation}
The above-mentioned results have been obtained at low energies,
i.e. in a semiclassical approximation. Because of this, the
problem arises how these results are changed in the case of high
energies? Or more precisely, how the results of
\cite{Jac1},\cite{Padm3}-- \cite{Padm12} are generalized in the
UV-limit?  It is obvious that, as in this case all the
thermodynamic characteristics become dependent on the deformation
parameter $\alpha$, all the corresponding results should be
modified (deformed) to meet the following requirements:
\\(a) to be clearly dependent on the deformation parameter
$\alpha$ at high energies;
\\
\\(b) to be duplicated, with high precision, at low energies
due to the suitable limiting transition;
\\
\\(c) Let us clear up what is meant by the adequate high energy
$\alpha$-deformation of Einstein's equations in similarity with
$\wp_{\varrho}$-deformation of the  General Relativity (formula
(2)\cite{Mark1}) that, as has been indicated in Section  3 of this
work,  is actually the $\alpha$-deformation.
\\ The problem may be more specific. As, according
to \cite{Jac1},\cite{Padm11},\cite{Padm12} and some other works,
gravitation is greatly determined by thermodynamics, and at high
energies the latter is a "deformation of the classical
thermodynamics". Here "high-energy deformation of thermodynamics"
is understood as some (meanwhile unknown) deformation of
thermodynamics in high energies. This theory is still unframed,
though several of its elements $T_{GUP},S_{GUP}$ and the like
\cite{acs}--\cite{Nou} are known already. It is interesting
whether gravitation at high energies (or what is the same, quantum
gravity or Planck scale)is being determined by the corresponding
deformed thermodynamics. The formulae (\ref{GT5}) and
(\ref{GT6.3}) are elements of the high-energy $\alpha$-deformation
in thermodynamics, a general pattern of which still remains to be
formed. Obviously, these formulae should be involved in the
general pattern giving better insight into the quantum gravity, as
they are applicable to black mini-holes (Planck black holes) which
may be a significant element of such a pattern. But what about
other elements of this pattern? How can we generalize the results
\cite{Jac1},\cite{Padm11},\cite{Padm12}when the IR-cutoff tends to
the UV-cutoff (formula (\ref{GT1}))? What are modifications of the
thermodynamic identity (\ref{GT8}) in a high-energy deformed
thermodynamics and how is it applied in high-energy (quantum)
gravity?
\\By authors opinion, the methods developed to solve the problem
of point (c) and elucidation of other above-mentioned problems may
form the basis for a new approach to solution of the quantum
gravity problem. And one of the keys to the {\bf quantum gravity}
problem is a better insight into the {\bf high-energy
thermodynamics}.
\section{$\alpha$--Representation of Einstein's Equations}
Let us consider $\alpha$-representation and high energy
$\alpha$-deformation of the Einstein's field equations for the
specific cases of horizon spaces (the point (c) of Section 4). In
so doing the results of the survey work \cite{Padm13} are used.
Then, specifically, for a static, spherically symmetric horizon in
space-time described by the metric
\begin{equation}\label{GT9}
ds^2 = -f(r) c^2 dt^2 + f^{-1}(r) dr^2 + r^2 d\Omega^2
\end{equation}
the horizon location will be given by simple zero of the function
$f(r)$, at $r=a$.
\\  It is known that for horizon spaces one can introduce
the temperature that can be identified with an analytic
continuation to imaginary time. In the case under consideration
(\cite{Padm13}, eq.(116))
\begin{equation}\label{GT10}
k_BT=\frac{\hbar cf'(a)}{4\pi}.
\end{equation}
Therewith, the condition $f(a)=0$ and $f'(a)\ne 0$ must be
fulfilled.
\\ Then at the horizon $r=a$ Einstein's field equations
\begin{equation}\label{GT11}
\frac{c^4}{G}\left[\frac{1}{ 2} f'(a)a - \frac{1}{2}\right] = 4\pi
P a^2
\end{equation}
may be written as the thermodynamic identity
(\ref{GT8})(\cite{Padm13} formula (119))
\begin{equation}\label{GT12}
   \underbrace{\frac{{{\hbar}} cf'(a)}{4\pi}}_{\displaystyle{k_BT}}
    \ \underbrace{\frac{c^3}{G{{\hbar}}}d\left( \frac{1}{ 4} 4\pi a^2 \right)}_{
    \displaystyle{dS}}
  \ \underbrace{-\ \frac{1}{2}\frac{c^4 da}{G}}_{
    \displaystyle{-dE}}
 = \underbrace{P d \left( \frac{4\pi}{ 3}  a^3 \right)  }_{
    \displaystyle{P\, dV}}
\end{equation}
where $P = T^{r}_{r}$ is the trace of the momentum-energy tensor
and radial pressure. In the last equation $da$ arises in the
infinitesimal consideration of Einstein's equations when studying
two horizons distinguished by this infinitesimal quantity $a$ and
$a+da$ (\cite{Padm13} formula (118)).
\\ Now we consider (\ref{GT12})  in new notation expressing $a$
in terms of the corresponding deformation parameter $\alpha$. Then
we have
\begin{equation}\label{GT13}
a=l_{min}\alpha^{-1/2}.
\end{equation}
Therefore,
\begin{equation}\label{GT14}
f'(a)=-2l^{-1}_{min}\alpha^{3/2}f'(\alpha).
\end{equation}
Substituting this into (\ref{GT11}) or into (\ref{GT12}), we
obtain in the considered case of Einstein's equations in the
"$\alpha$--representation" the following:
\begin{equation}\label{GT16}
\frac{c^{4}}{G}(-\alpha f'(\alpha)-\frac{1}{2})=4\pi
P\alpha^{-1}l^{2}_{min}.
\end{equation}
Multiplying the left- and right-hand sides of the last equation by
$\alpha$, we get
\begin{equation}\label{GT16.1}
\frac{c^{4}}{G}(-\alpha^{2}f'(\alpha)-\frac{1}{2}\alpha)=4\pi
Pl^{2}_{min}.
\end{equation}
But since usually $l_{min}\sim l_{p}$ (that is just the case if
the Generalized Uncertainty Principle (GUP) is satisfied), we have
$l^{2}_{min}\sim l^{2}_{p}=G\hbar/c^{3}$. When selecting a system
of units, where $\hbar=c=1$, we arrive at $l_{min}\sim l_{p}=\surd
G$, and then (\ref{GT16}) is of the form
\begin{equation}\label{GT16.A}
-\alpha^{2}f'(\alpha)-\frac{1}{2}\alpha=4\pi P\vartheta^{2}G^{2},
\end{equation}
where $\vartheta=l_{min}/l_{p}$. L.h.s. of (\ref{GT16.A}) is
dependent on $\alpha$. Because of this, r.h.s. of (\ref{GT16.A})
must be dependent on $\alpha$ as well, i. e. $P=P(\alpha)$.
\begin{center}
{\bf Analysis of $\alpha$-Representation of Einstein's
Equations}
\end{center}
Now let us get back to (\ref{GT12}). In \cite{Padm13} the
low-energy case has been considered, for which (\cite{Padm13}
formula (120))
\begin{equation}\label{GT17.A}
 S=\frac{1}{ 4l_p^2} (4\pi a^2) = \frac{1}{ 4} \frac{A_H}{ l_p^2}; \quad E=\frac{c^4}{ 2G} a
    =\frac{c^4}{G}\left( \frac{A_H}{ 16 \pi}\right)^{1/2},
\end{equation}
where $A_H$ is the horizon area. In our notation (\ref{GT17.A})
may be rewritten as
\begin{equation}\label{GT17.A1}
 S= \frac{1}{4}\pi\alpha^{-1}; \quad E=\frac{c^4}{2G} a
 =\frac{c^4}{G}\left( \frac{A_H}{ 16 \pi}\right)^{1/2}=\frac{\vartheta}{2\surd G}\alpha^{1/2}.
\end{equation}
We proceed to two entirely different cases: low energy (LE) case
and high energy (HE) case. In our notation these are respectively
given by
\begin{center}
A)$\alpha\rightarrow 0$ (LE), B)$\alpha\rightarrow 1/4$ (HE),
\\C)$\alpha$ complies with the familiar scales and energies.
\end{center}
The case of C) is of no particular importance as it may be
considered within the scope of the conventional General
Relativity.
\\Indeed, in point A)$\alpha\rightarrow 0$ is not actually an exact
limit as a real scale of the Universe (Infrared (IR)-cutoff
$l_{max}\approx 10^{28}cm$), and then
\begin{center}
$\alpha_{min}\sim l_{p}^{2}/l^{2}_{max}\approx 10^{-122}$.
\end{center}
In this way A) is replaced by A1)$\alpha\rightarrow \alpha_{min}$.
In any case at low energies the second term in the left-hand side
(\ref{GT16.A}) may be neglected in the  infrared limit.
Consequently, at low energies (\ref{GT16.A}) is written as
\begin{equation}\label{GT16.LE}
-\alpha^{2}f'(\alpha)=4\pi P(\alpha)\vartheta^{2}G^{2}.
\end{equation}
Solution of the corresponding Einstein equation – finding of the
function $f(\alpha)=f[P(\alpha)]$ satisfying(\ref{GT16.LE}). In
this case formulae (\ref{GT17.A}) are valid as at low energies a
semiclassical approximation is true. But from (\ref{GT16.LE})it
follows that
\begin{equation}\label{GT16.solv}
f(\alpha)=-4\pi \vartheta^{2}G^{2}\int
\frac{P(\alpha)}{\alpha^{2}}d\alpha.
\end{equation}
On the contrary, knowing $f(\alpha)$, we can obtain
$P(\alpha)=T^{r}_{r}.$
\\ But it is noteworthy that, when studying the infrared modified gravity
\cite{Patil},\cite{Park1},\cite{Rub}, we have to make corrections
for the considerations of point A1).
\section{Possible High Energy $\alpha$-Deformation of General Relativity}
Let us consider the high-energy case B). Here two variants are
possible.
\\
\\{\bf I. First variant}.
\\ In this case it is assumed that in the high-energy
(Ultraviolet (UV))limit the thermodynamic identity (\ref{GT12})(or
that is the same (\ref{GT8})is retained but now all the quantities
involved in this identity become $\alpha$-deformed. This means
that they appear in the $\alpha$-representation with quantum
corrections and are considered at high values of the parameter
$\alpha$, i.e. at $\alpha$ close to 1/4. In particular, the
temperature $T$ from equation (\ref{GT12}) is changed by $T_{GUP}$
(\ref{GT5}), the entropy $S$ from the same equation given by
semiclassical formula (\ref{GT17.A}) is changed by $S_{GUP}$
(\ref{GT6.3}), and so forth:
\begin{center}
$E\mapsto E_{GUP}, V\mapsto V_{GUP}$.
\end{center}
Then the high-energy $\alpha$-deformation of equation (\ref{GT12})
takes the form
\begin{equation}\label{GT8.GUP}
k_{B}T_{GUP}(\alpha)dS_{GUP}(\alpha)-dE_{GUP}(\alpha)=P(\alpha)dV_{GUP}(\alpha).
\end{equation}
Substituting into (\ref{GT8.GUP}) the corresponding quantities
\\$T_{GUP}(\alpha),S_{GUP}(\alpha),E_{GUP}(\alpha),V_{GUP}(\alpha),P(\alpha)$
and expanding them into a Laurent series in terms of $\alpha$,
close to high values of $\alpha$, specifically close to
$\alpha=1/4$, we can derive a solution for the high energy
$\alpha$-deformation of general relativity (\ref{GT8.GUP}) as a
function of $P(\alpha)$. As this takes place, provided at high
energies the generalization of (\ref{GT12}) to (\ref{GT8.GUP})is
possible, we can have the high-energy $\alpha$-deformation of the
metric. Actually, as from (\ref{GT12}) it follows that
\begin{equation}\label{GT8.GUP1}
f'(a)=\frac{4\pi k_{B}}{\hbar c}T=4\pi k_{B}T
\end{equation}
(considering that we have assumed $\hbar=c=1$), we get
\begin{equation}\label{GT8.GUP2}
f'_{GUP}(a)=4\pi k_{B}T_{GUP}(\alpha).
\end{equation}
L.h.s. of (\ref{GT8.GUP2}) is directly obtained in the
$\alpha$-representation. This means that, when $f'\sim T$, we have
$f'_{GUP}\sim T_{GUP}$ with the same factor of proportionality. In
this case the function $f_{GUP}$ determining the high-energy
$\alpha$-deformation of the spherically symmetric metric may be in
fact derived by the expansion of $T_{GUP}$, that is known from
(\ref{GT5}), into a Laurent series in terms of  $\alpha$ close to
high values of $\alpha$ (specifically close to $\alpha=1/4$), and
by the subsequent integration.
\\ It might be well to remark on the following.
\\
\\{\bf 6.1} As on going to high energies we use (GUP),
$\vartheta$ from equation (\ref{GT16.A})is expressed in terms of
$\alpha^{\prime}$--dimensionless constant from GUP
(\ref{GUP1}),(\ref{GT5}):$\vartheta=2\alpha^{\prime}.$
\\
\\{\bf 6.2} Of course, in all the formulae including $l_{p}$
this quantity must be changed by $G^{1/2}$ and hence $l_{min}$
by $\vartheta G^{1/2}=2\alpha^{\prime} G^{1/2}.$
\\
\\{\bf 6.3} As noted in the end of subsection 6.1,
and in this case also knowing all the high-energy deformed
quantities
$T_{GUP}(\alpha),S_{GUP}(\alpha),E_{GUP}(\alpha),V_{GUP}(\alpha)$,
we can find $P(\alpha)$ at $\alpha$ close to 1/4.
\\
\\{\bf 6.4} Here it is implicitly understood that the Ultraviolet
limit of Einstein's  equations is independent of the starting
horizon space. This assumption is quite reasonable. Because of
this, we use  the well-known formulae for the modification of
thermodynamics and statistical mechanics of black holes in the
presence of GUP \cite{acs}--\cite{Nou}
\\
\\{\bf 6.5} The use of the thermodynamic identity
(\ref{GT8.GUP}) for the description of the high energy deformation
in General Relativity implies that on going to the UV-limit of
Einstein's equations for horizon spaces in the thermodynamic
representation (consideration) we are trying to remain within the
scope of {\bf equilibrium statistical mechanics} \cite{Balesku1}
({\bf equilibrium thermodynamics}) \cite{Bazarov}. However, such
an assumption seems to be too strong. But some grounds to think so
may be found as well. Among other things, of interest is the
result from \cite{acs} that GUP may prevent black holes from their
total evaporation. In this case the Planck's remnants of black
holes will be stable, and when they are considered, in some
approximation the {\bf equilibrium thermodynamics} should be
valid. At the same time, by author's opinion these arguments are
rather weak to think that the quantum gravitational effects in
this context have been described only within the scope of {\bf
equilibrium thermodynamics} \cite{Bazarov}.
\\
\\{\bf II. Second variant}.
\\ According to the remark of {\bf 6.5},
it is assumed that the interpretation of Einstein's equations as a
thermodynamic identity (\ref{GT12}) is not retained on going to
high energies (UV--limit), i.e. at $\alpha\rightarrow 1/4$, and
the situation is adequately described exclusively by {\bf
non-equilibrium thermodynamics} \cite{Bazarov},\cite{Gyarm}.
Naturally, the question arises: which of the additional terms
introduced in (\ref{GT12}) at high energies may be leading to such
a description?
\\In the \cite{shalyt-aip},\cite{shalyt-entropy2} it has been shown that in case the cosmological
term $\Lambda$ is a dynamic quantity, it is small at low energies
and may be sufficiently large at high energies. In the right-hand
side of (\ref{GT16.A}) in the $\alpha$--representation the
additional term $G(\Lambda(\alpha))$ is introduced:
 \begin{equation}\label{GT16.B}
 -\alpha^{2}f'(\alpha)-\frac{1}{2}\alpha=4\pi P(\alpha)\vartheta^{2}G^{2}-G\Lambda(\alpha),
 \end{equation}
 where $\Lambda(\alpha)$ is the cosmological term depending from $\alpha$.
  Then its inclusion
in the low-energy case (\ref{GT11})(or in the $\alpha$
-representation (\ref{GT16.A})) has actually no effect on the
thermodynamic identity (\ref{GT12}) validity, and consideration
within the scope of equilibrium thermodynamics still holds true.
It is well known that this is not the case at high energies as the
$\Lambda$-term may contribute significantly to make the "process"
non-equilibrium in the end \cite{Bazarov},\cite{Gyarm}.
\\ Is this the only cause for violation of the thermodynamic
identity (\ref{GT12}) as an interpretation of the high-energy
generalization of Einstein's equations? Further investigations are
required to answer this question.
\section{Deformed Gravity and Dark Energy Problem}
Let us revert to Section 3 and to the above-mentioned work
\cite{Mark1} from the viewpoint of item II of the previous
Section. It is obvious that in model (\ref{Mark5})(\cite{Mark1},
formula (2)) for $\wp_{\varrho}$-deformation of the  General
Relativity, since the right-hand side is dependent on the
parameter $\wp_{\varrho}$, the left-hand side is also dependent on
this parameter, i.e. (\ref{Mark5}) may be written as
\begin{equation}\label{DE1}
R^{\nu}_{\mu}(\wp_{\varrho})-\frac{1}{2}R\delta^{\nu}_{\mu}(\wp_{\varrho})=\frac{8\pi
G}{c^{4}}T^{\nu}_{\mu}(1-\wp_{\varrho}^{2})^{n}-\Lambda\wp_{\varrho}^{2n}\delta^{\nu}_{\mu},
\end{equation}
where the dependence of the left side on $\wp_{\varrho}$ comes to
naught when $\wp_{\varrho}\ll 1$. Otherwise, it should be taken
into account.
\\ But, according to (\ref{Mark8}), $\wp_{\varrho}\sim \alpha^{2}$
and hence in fact (\ref{DE1}) is the $\alpha$-deformation of
Einstein's Equations
\begin{equation}\label{DE2}
R^{\nu}_{\mu}(\alpha)-\frac{1}{2}R\delta^{\nu}_{\mu}(\alpha)=\frac{8\pi
G}{c^{4}}T^{\nu}_{\mu}(\alpha)-\Lambda(\alpha)\delta^{\nu}_{\mu}.
\end{equation}
It is clear that $\alpha$-deformation (\ref{DE2}) of the Einstein
Equations is similar to the $\alpha$-deformation (\ref{GT16.B})
from the previous Section. At the same time, they are
significantly different: the first is purely heuristic, whereas
the second has been obtained using the high-energy
$\alpha$-deformation of the General Relativity in the case when it
permits a thermodynamic interpretation (\ref{GT8}),(\ref{GT12}).
Of course, we consider the horizon spaces only and also the cases
when the Gravitation field equations on the horizon may be
represented in the form of a thermodynamic identity (\ref{GT8}).
Now the number of such cases is minor, all of them being mentioned
with the corresponding references in Section 5.2 of \cite{Padm13}.
\\ In this way, proceeding from the results of Sections 5 and 6 ,
it may be assumed that  $\alpha$-deformation
(\ref{DE2}) of the General Relativity is often the case. And the de
Sitter Universe is the case on condition that
\begin{equation}\label{DE3}
\lim\limits_{\alpha\rightarrow 1/4}T^{\nu}_{\mu}(\alpha)
\rightarrow 0 .
\end{equation}
The problem is, what is the dependence $\Lambda(\alpha)$ on
$\alpha$, to give the adequate value of $\Lambda(\alpha)$   within
the scope of a dynamic model $\Lambda=\Lambda(t)$
\cite{varyingL}--\cite {Var-CC2} at the present time. As by
formula (\ref{Mark8} from Section 3 $\wp_{\varrho}\sim
\alpha^{2}$, the main equation (\ref{Mark5}) of \cite{Mark1}
suggests that in this case we have
\begin{equation}\label{DE4}
\Lambda(\wp_{\varrho}) \sim  \alpha^{4n} \Lambda.
\end{equation}
And, since in this case $n\geq 1/2$, a "minimal" dependence
$\Lambda(\alpha)$  on  $\alpha$ will be given by
\begin{equation}\label{DE5}
\Lambda(\alpha) \sim \alpha^{2} \Lambda.
\end{equation}
However, as shown in \cite{shalyt-aip},\cite{shalyt-entropy2},
within the scope of the holographic principle
\cite{Hooft1}--\cite{Bou3} we actually have
\begin{equation}\label{DE6}
\Lambda(\alpha) \sim \alpha \Lambda,
\end{equation}
where in the right side of (\ref{DE6})  $\Lambda$  is understood
as a cosmological constant  at Planck's scales
$\Lambda=\Lambda_{p}$.
\\ Let us consider the calculations from
\cite{shalyt-aip},\cite{shalyt-entropy2} in greater detail.
\\We begin with the Schwarzschild black holes, whose
 semiclassical entropy is given by
\begin{equation}\label{D1.1}
S = \pi {R_{Sch}^2}/ l_p^2=\pi {R_{Sch}^2}
m_p^2=\pi\alpha_{R_{Sch}}^{-1},
\end{equation}
with the assumption that in the formula for $\alpha$ $R_{Sch}=x$
is the measuring scale and $l_p = 1/m_p$. Here $R_{Sch}$ is the
adequate Schwarzschild radius, and $\alpha_{R_{Sch}}$ is the value
of $\alpha$ associated with this radius. Then, as it has been
pointed out in \cite{Bal}, in case the Fischler- Susskind cosmic
holographic conjecture \cite{Sussk1} is valid, the entropy of the
Universe is limited by its "surface"  measured in Planck units
\cite{Bal}:
\begin{equation}\label{D2}
S \leq \frac{A}{4} m_p^2,
\end{equation}
where the surface area $A = 4\pi R^2$ is defined in terms of the
apparent (Hubble) horizon
\begin{equation}\label{D3}
R = \frac{1}{\sqrt{H^2+k/a^2}},
\end{equation}
with curvature $k$  and scale $a$ factors.
\\ Again, interpreting $R$ from (\ref{D3}) as a measuring scale,
we directly obtain(\ref{D2}) in terms of $\alpha$:
\begin{equation}\label{D4}
S \leq \pi\alpha_{R}^{-1},
\end{equation}
where $\alpha_{R}=l^{2}_{p}/R^{2}$. Therefore, the average entropy
density may be found as
\begin{equation}\label{D5}
\frac{S}{V}\leq \frac{\pi \alpha_{R}^{-1}}{V}.
\end{equation}
Using further the reasoning line of \cite{Bal} based on the
results of the  holographic thermodynamics, we can relate the
entropy and energy of a holographic system
\cite{Jac1},\cite{Cai1}. Similarly, in terms of the $\alpha$
parameter one can easily estimate the upper limit for the energy
density of the Universe (denoted here by $\rho_{hol}$):
\begin{equation}\label{D6}
\rho_{hol} \leq \frac{3}{8 \pi R^2} m_p^2 = \frac{3}{8
\pi}\alpha_{R} m_p^4,
\end{equation}
that is drastically differing from the one obtained with
well-known QFT
\begin{equation}\label{D7}
\rho^{QFT}\sim m_p^4.
\end{equation}
Here by $\rho^{QFT}$ we denote the energy vacuum density
calculated from well-known QFT (without UV cutoff) \cite{Zel1}.
Obviously, as $\alpha_{R}$ for $R$ determined by (\ref{D3}) is
very small, actually approximating zero, $\rho_{hol}$ is by
several orders of magnitude smaller than the value expected in QFT
-- $\rho^{QFT}$.
\\In fact, the upper limit of the right-hand side of (\ref{D6})
is attainable, as it has been indicated in \cite{Bal}. The
"overestimation" value of $r$ for the energy density $\rho^{QFT}$,
compared to $\rho_{hol}$, may be determined as
\begin{equation}\label{D8}
r =\frac{\rho^{QFT}}{\rho_{hol}}=\frac{8 \pi}{3}{\bf
\alpha_{R}^{-1}}
 =\frac{8 \pi}{3} \frac{R^2}{l_p^2}
 =\frac{8 \pi}{3} \frac{S}{S_p},
\end{equation}
where $S_p$ is the entropy of the Plank mass and length for the
Schwarzschild black hole. It is clear that due to smallness of
$\alpha_{R}$ the value of $\alpha_{R}^{-1}$ is on the contrary too
large. It may be easily calculated (e.g., see \cite{Bal})
\begin{equation}\label{D9}
r = 5.44\times 10^{122}
\end{equation}
in a good agreement with the astrophysical data.
\\ Naturally, on the assumption that the vacuum energy density
$\rho_{vac}$ is involved in $\rho$ as a term
\begin{equation}\label{vac1}
\rho = \rho_M + \rho_{vac},
\end{equation}
where $\rho_M$ - average matter  density, in case of $\rho_{vac}$
we can arrive to the same upper limit (right-hand side of the
formula (\ref{D6})) as for $\rho$.
\\ As the density of the vacuum energy $\rho_{vac}$
is nothing else  but $\Lambda$: $\Lambda \sim \rho_{vac}$,
according to the above calculations, we have
\begin{equation}\label{DE7}
\Lambda(\alpha) \sim \alpha \Lambda_{p}.
\end{equation}
And this explains the fact that in modern period the experimental
value $\Lambda=\Lambda_{exper}$ is lower than that derived in
conventional QFT  by the cut-off method at Planck's scales
\cite{Zel1}, \cite{Wein1} by a factor of $\sim 10^{-122}$, because
the corresponding value of $\alpha$  is given by
\begin{equation}\label{DE8}
\alpha \sim \frac{l^{2}_{p}}{R^{2}}\approx 10^{-122},
\end{equation}
where $R=10^{28}cm$ -- radius of the observable part of the Universe.
\\ As for the derivation of (\ref{DE6}) a semiclassical approximation has been used,
in(\ref{DE6}) in the right side the factor for $\Lambda$ is
actually a series in $\alpha$, i.e. in the general case(\ref{DE6})
it takes the form
\begin{equation}\label{DE9}
\Lambda(\alpha) \sim (\alpha +\xi_{1}\alpha^{2}+...)\Lambda_{p}
\end{equation}
and this is analogous to the series expansion
\begin{equation}\label{DE10}
\rho_{\rm vac}={\frac{1}{l_p^4}}+
{\frac{1}{l_p^4}\left(\frac{l_p}{l_\Lambda}\right)^2}
+{\frac{1}{l_p^4}\left(\frac{l_p}{l_\Lambda}\right)^4} + ...,
\end{equation}
in (\cite{Padm1},formula (33)),(\cite{Padm2},formula (12)).
\\Note that the holographic principle is valid for horizon spaces as it has been found
in \cite{Padm3}--\cite{Padm11},\cite{Padm13}, since in this case
Einstein's equations may be derived from the surface term of the
Lagrangian because it contains the same information  as the bulk
term.
\\As demonstrated by the above calculations, to meet the experimental data, the heuristic model
\cite{Mark1} must be corrected, since within the scope of the
$\alpha$-deformation a correct value of $\Lambda_{exper}$ for the
modern period is given by the formula of (\ref{DE6}) rather than
by the formula(\ref{DE5}) that is directly inferred from this
model.
\\Also, it should be noted that within a dynamic model for $\Lambda$ the
Uncertainty Principle derived in \cite{Min1}--\cite{Min4} for the
pair $(\Lambda, V)$, where  $V$ -- "four-dimensional" volume, has
been extended in \cite{shalyt-aip},\cite{shalyt-entropy2} to the
Generalized Uncertainty Principle, where at a qualitative level
the drastic distinctions between $\Lambda_{exper}$ and $\Lambda$
calculated with the use of QFT \cite{Zel1}, \cite{Wein1} are
explained.
\section{Conclusion}
The results obtained in Sections 6 and 7 enable framing of the following hypothesis:
\\{\bf a dynamic cosmological term is a measure of deviation from the thermodynamic
identity (the first law of thermodynamics) of the high-energy
(Planck's) deformation of Einstein equations for horizon spaces in
their thermodynamic interpretation.}
\\ The dynamic cosmological term correlates well with inflation models
\cite{Kolb} as the latter require a very high $\Lambda$ at the
early  stages of  the Universe, and this is distinct from
$\Lambda=\Lambda_{exper}$  in the modern period.
 Of great interest is the recent work \cite{Polyak}, where a mechanism
of the vacuum energy decay in the de Sitter space is established
to support a dynamic nature of $\Lambda$.
\\ This work is a  step to the incorporation of
deformation parameters involved in a quantum field theory at
Planck's scales into the high-energy deformation of the General
Relativity (GR). The corresponding calculations with the adequate
interpretation must follow next. It is interesting to consider the
high energy $\alpha$-deformation of GR  in a more general case.
The problem is how far a thermodynamic interpretation of
Einstein's equations may be extended? We should remember that, as
all the deformations considered involve a minimal length at the
Planck level $l_{min}\sim l_{p}$, a minimal volume should also be
the case $V_{min}\sim V_{p}=l^{3}_{p}$. And this is of particular
importance for high energy thermodynamics (some indications to
this fact have been demonstrated in
\cite{shalyt-aip},\cite{shalyt-entropy2}.
\\ Besides, in this paper we have treated QFT
with a minimal length, i.e. with the UV-cutoff. Consideration of
QFT with a minimal momentum (or IR-cutoff) \cite{Kim1}
necessitates an adequate extension of the $\alpha$-deformation in QFT
with the introduction of new parameters significant in the
IR-limit. It seems that some hints to a nature of such deformation
may be found in the works devoted to the infrared modification
of gravity \cite{Patil}--\cite{Rub}.


\begin{thebibliography}{1}
\bibitem{Planck1}
Klinkhamer F.R.Fundamental length scale of quantum spacetime foam.
\emph{JETPLett.} {\bf 2007}, {\em 86}, 2167-2180.
%
%
\bibitem{Planck2}
Amelino-Camelia G.;  Smolin L. Prospects for constraining quantum
gravity dispersion with near term observations. \emph{Phys.Rev.D}
{\bf 2009}, {\em 80}, 084017;  Gubitosi G. et al. A Constraint on
Planck-scale Modifications to Electrodynamics with CMB
polarization data. \emph{JCAP} {\bf 2009}, {\em 0908}, 021;
Amelino-Camelia G. Building a case for a Planck-scale-deformed
boost action: the Planck-scale particle-localization limit.
\emph{Int.J.Mod.Phys.D} {\bf 2005}, {\em 14}, 2167-2180.
%
%
\bibitem{Planck3}
Hossenfelder S. et al. Signatures in the Planck Regime. \emph
{Phys. Lett.B} {\bf 2003}, {\em 575}, 85-99; Hossenfelder S.,
Running coupling with minimal length \emph{Phys.Rev.D} {\bf 2004}
{\em 70}, 105003; Hossenfelder S., Self-consistency in theories
with a minimal length, \emph{Class. Quant. Grav.} {\bf 2006}, {\em
23}, 1815-1821.
%
%
\bibitem{Fadd}
Faddeev, L., Mathematical View on Evolution of Physics. {\em
Priroda} {\bf 1989}, {\em 5}, 11--18.
%
%
\bibitem{Sci1}
Sciama, D.W., 1984, Proc.R.Soc.Lond. A 394, 1, Bowyer S.,
KorpelaE.J., Edelstein J. , Lampton M., Morales C., Perez-Mercader
J., Gomez J.F., Trapero J., 1999, ApJ 526, 10, Turner, M.S., 1991,
Physica scripta. vol T36, 167.
%
%
\bibitem{Dar1}
Perlmutter, S.  et al. Measurements of Omega and Lambda from 42
high redshift supernovae. {\em Astrophys. J} {\bf 1999}, {\em
517}, 565--586; Riess A. G. et al. Observational evidence from
supernovae for an accelerating universe and a cosmological
constant. {\em Astron. J} {\bf 1998}, {\em 116}, 1009--1038; Riess
A. G. et al. BV RI light curves for 22 type Ia supernovae. {\em
Astron. J} {\bf 1999}, {\em 117}, 707--724; Sahni, V. and
Starobinsky, A. A. The Case for a positive cosmological Lambda
term. \emph{Int. J. Mod. Phys. D} {\bf 2000} {\em 9}, 373--397;
Carroll, S. M. The Cosmological constant. \emph{Living Rev. Rel}
{\bf 2001}, {\em 4}, 1--50; Padmanabhan, T. Cosmological constant:
The Weight of the vacuum. \emph{Phys. Rept} {\bf 2003}, {\em 380},
235--320; Padmanabhan, T. Dark Energy: the Cosmological Challenge
of the Millennium. \emph{Current Science} {\bf 88}, 1057--1071
(2005); Peebles, P. J. E. and Ratra, B. The Cosmological constant
and dark energy. \emph{Rev. Mod. Phys} {\bf 2003}, {\em 75},
559--606.
%
%
\bibitem{Hoop}
 Dan Hooper, TASI 2008 Lectures on Dark Matter, \emph{ArXiv:
 0901.4090}
%
%
\bibitem{Quint1}
Ratra,B.  and Peebles, J. Cosmological Consequences of a Rolling
Homogeneous Scalar Field. \emph{Phys. Rev. D} {\bf 1988}, {\em
37}, 3406--3422; Caldwell,R. R., Dave, R. and Steinhardt, P. J.
 Cosmological imprint of an energy component with general equation
of state. \emph{Phys. Rev. Lett} {\bf 1998}, {\em 80}, 1582--1585.
%
%
\bibitem{K1}
Armendariz-Picaon, C., Damour,T.  and V. Mukhanov, V.
 k - inflation. \emph{Phys. Lett. B} {\bf 1999}, {\em 458}, 209--218 ;
J. Garriga, and V. Mukhanov, Perturbations in k-inflation.
 \emph{Phys. Lett. B} {\bf 1999} {\em 458}, 219--225 (1999).
%
%
\bibitem{Tach1}
Padmanabhan, T. Accelerated expansion of the universe driven by
tachyonic matter. \emph{Phys. Rev. D} {\bf 2002} {\bf 66}, 021301;
Bagla, J. S.  Jassal,H. K. and Padmanabhan, T. Cosmology with
tachyon field as dark energy. \emph{Phys. Rev. D}
 {\bf 2003},{\em 67}, 063504 ; Abramo, L. R. W., and
Finelli, F. Cosmological dynamics of the tachyon with an inverse
power-law potential. \emph{Phys. Lett. B}  {\bf 2003},{\em 575},
165--171; Aguirregabiria J. M. and Lazkoz, R. Tracking solutions
in tachyon cosmology. \emph{Phys. Rev. D} {\bf 2004}, {\em 69},
123502 ; Guo, Z. K. and Zhang, Y. Z. Cosmological scaling
solutions of the tachyon with multiple inverse square potentials.
\emph{JCAP} {\bf 2004} {\em 0408}, 010;  Copeland, E. J. et al.
 What is needed of a tachyon if it is to be the dark energy?
 \emph{Phys. Rev. D} {\bf 2005} {\em 71}, 043003.
%
%
\bibitem{Phant1}
Sahni, V. and  Shtanov, Y. Brane world models of dark energy.
 \emph{JCAP} {\bf 2003}, {\em 0311}, 014;
  Elizalde, E., Nojiri, S.,  and Odintsov, S. D.
 Late-time cosmology in (phantom) scalar-tensor theory:
 Dark energy and the cosmic speed-up.
 \emph{Phys. Rev. D} {\bf 2004}, {\em 70}, 043539.
%
%
 \bibitem{Gliner}
Gliner, E. B. \emph{ZHETF} {\bf 1965} {\em 49}, 542--549.
%
%
\bibitem{Zel1}
Zel'dovich,Y.B. \emph{Sov.Phys.Uspehi} {\bf 1968} {\em 11},
381--393.
%
%
\bibitem{Wein1}
Weinberg, S. The Cosmological Constant Problem. \emph{Rev. Mod.
Phys} {\bf 1989} {\em 61}, 1--23.
%
%
\bibitem{Wald}
Wald, Robert. M. General Relativity. \emph{The University Chicago
Press}. Chicago and London 1984, 491 p.p.
%
%
\bibitem{varyingL}
O. Bertolami, N. Cim. B {\bf 93}, 36 (1986); J.C. Carvalho, J.A.S
Lima and I. Waga, Phys. Rev. D {\bf 46}, 2404 (1992); L.P.
Chimento and D. Pavon, Gen. Rel. Grav. {\bf 30}, 643 (1998); T.
Harco and M.K. Mak, Gen. Rel. Grav. {\bf 31} 849 (1999); S.
Carneiro, arxiv:gr-qc/0307114
%
%
\bibitem{Var-CC1}
R. Aldrovandi, J. P. Beltran Almeida, J. G. Pereira,Time-Varying
Cosmological Term: Emergence and Fate of a FRW Comments.
\emph{Grav.Cosmol.} 11 (2005) 277-283
%
%
\bibitem{Var-CC2}
Richard T Hammond, Terry Pilling, Dark Entropy,arXiv:0806.1277
%
%
\bibitem{Ven1}
Veneziano,G. A Stringy Nature Needs Just Two Constants
\emph{Europhys.Lett} {\bf 1986}, {\em 2}, 199--211; Amati, D.;
Ciafaloni, M., and Veneziano,G. Can Space-Time Be Probed Below the
String Size? \emph{Phys.Lett.B} {\bf 1989}, {\bf 216}, 41--47;
E.Witten, \emph{Phys.Today} {\bf 1996}, {\em 49}, 24--28.
%
%
\bibitem{GUPg1}
~Adler,R.~J.; ~Santiago,D.~I.
 On gravity and the uncertainty principle. \emph{Mod. Phys. Lett.
A} {\bf 1999}, {\em 14}, 1371--1378.
%
\bibitem{GUPg2}
~Scardigli,F. Generalized uncertainty principle in quantum gravity
from micro - black hole Gedanken experiment. \emph{Phys. Lett. B}
{\bf 1999}, {\em 452}, 39--44; ~Bambi,C. A Revision of the
Generalized Uncertainty Principle. \emph{Class. Quant. Grav}
 {\bf 2008}, {\em 25}, 105003.
%
%
\bibitem{Gar1}
Garay,L. Quantum gravity and minimum length.
\emph{Int.J.Mod.Phys.A} {\bf 1995}, {\em 10}, 145--166.
%
%
\bibitem{Ahl1}
Ahluwalia,D.V. Wave particle duality at the Planck scale: Freezing
of neutrino oscillations. \emph{Phys.Lett} {\bf 2000}, {\em A275},
31--35; Ahluwalia,D.V. \emph{Mod.Phys.Lett} {\bf 2002}, Interface
of gravitational and quantum realms. {\em A17}, 1135--1146.
%
%
\bibitem{Magg}
Maggiore,M. A Generalized uncertainty principle in quantum
gravity. \emph{Phys.Lett} {\bf 1993}, {\em B304}, 65--69.
%
%
\bibitem{Magg1}
Maggiore,M. The Algebraic structure of the generalized uncertainty
principle. \emph{Phys.Lett.B} {\bf 1993}, {\em 319}, 83--86.
%
%
\bibitem{Kempf}
Kempf,A.; Mangano,G.; Mann,R.B. Hilbert space representation of
the minimal length uncertainty relation. \emph{Phys.Rev.D} {\bf
1995}, {\em 52}, 1108--1118.
%
%
\bibitem{shalyt1}
Shalyt-Margolin, A.E.; Suarez, J.G. Quantum mechanics of the early
universe and its limiting transition. \emph{gr-qc/0302119}, 16pp.
%
%
\bibitem{shalyt2}
Shalyt-Margolin, A.E.; Suarez, J.G. Quantum mechanics at Planck's
scale and density matrix. \emph{Intern. Journ. Mod. Phys D} {\bf
2003}, {\em 12}, 1265--1278.
%
%
\bibitem{shalyt3}
Shalyt-Margolin, A.E.; Tregubovich, A.Ya. Deformed density matrix
and generalized uncertainty relation in thermodynamics.
 \emph{Mod. Phys.Lett. A} {\bf 2004}, {\em 19}, 71--82.
%
%
\bibitem{shalyt4}
Shalyt-Margolin, A.E. Nonunitary and unitary transitions in
generalized quantum mechanics, new small parameter and information
problem solving. \emph{Mod. Phys. Lett. A} {\bf 2004}, {\em 19},
391--404.
%
%
\bibitem{shalyt5}
Shalyt-Margolin, A.E. Pure states, mixed states and Hawking
problem in generalized quantum mechanics. \emph{Mod. Phys. Lett.
A} {\bf 2004}, {\em 19}, 2037--2045.
%
%
\bibitem{shalyt6}
Shalyt-Margolin, A.E. The Universe as a nonuniform lattice in
finite volume hypercube. I. Fundamental definitions and particular
features\emph{Intern. Journ. Mod.Phys D} {\bf 2004}, {\em 13},
853-- 864.
%
%
\bibitem{shalyt7}
Shalyt-Margolin, A.E. The Universe as a nonuniform lattice in the
finite-dimensional hypercube. II. Simple cases of symmetry
breakdown and restoration. \emph{Intern.Journ.Mod.Phys.A} {\bf
2005}, {\em 20}, 4951--4964.
%
%
\bibitem{shalyt8}
Shalyt-Margolin, A.E.; Strazhev,V.I. The Density Matrix
Deformation in Quantum and Statistical Mechanics in Early
Universe. In \emph{Proc. Sixth International Symposium "Frontiers
of Fundamental and Computational Physics"},edited by B.G. Sidharth
at al. Springer,2006, pp.131--134.
%
%
\bibitem{shalyt9}
Shalyt-Margolin, A.E. The Density matrix deformation in physics of
the early universe and some of its implications. In \emph{Quantum
Cosmology Research Trends},edited by A. Reimer, Horizons in World
Physics. {\bf 246}, Nova Science Publishers, Inc., Hauppauge,
NY,2005, pp. 49--91.
%
%
\bibitem{shalyt10}
Shalyt-Margolin, A.E. Deformed density matrix and quantum entropy
of the black hole. \emph{Entropy} {\bf 2006}, {\em 8}, 31--43.
%
%
\bibitem{shalyt13}
Shalyt-Margolin, A.E. Entropy in the Present and Early Universe.
\emph{Symmetry} {\bf 2007}, {\em 18
}, 299--320.
%
%
\bibitem{Mark1}
Markov, M.À. Limiting density of matter as the universal law of
nature. \emph{Pis'ma v ZHETF} {\bf 1982}, {\em 36 }, 214--216.
%
%
\bibitem{shalyt-aip}
Shalyt-Margolin,A.E. Entropy in the Present and Early Universe and
Vacuum Energy, \emph{AIP Conference Proceedings},  {\bf 2010},
{\em 1205}, 160--167.
%
%
\bibitem{shalyt-entropy2}
Shalyt-Margolin,A.E. Entropy In The Present And Early Universe:
New Small Parameters And Dark Energy Problem \emph{Entropy} {\bf
2010}, {\em 1205},  932-952
%
%
\bibitem{Kim2}
Kim,Yong-Wan; Lee,Hyung Won; Myung, Yun Soo. Entropy bound of
local quantum field theory with generalized uncertainty principle.
\emph{Phys.Lett.B} {\bf 2009}, {\em 673}, 293-296.
%
%
\bibitem{Padm1}
Padmanabhan,T. Darker side of the universe ... and the crying need
for some bright ideas! \emph{Proceedings of the 29th International
Cosmic Ray Conference}, Pune, India,2005; pp. 47--62.
%
%
\bibitem{Padm2}
Padmanabhan,T. Dark Energy: Mystery of the Millennium.
 \emph{Paris 2005, Albert Einstein's century }, AIP Conference
Proceedings 861, American Institute of Physics, New York, 2006;
pp. 858--866.
%
%
\bibitem{Padm3}
Padmanabhan,T. A New perspective on gravity and the dynamics of
spacetime. \emph{Int.Jorn.Mod.Phys} {\bf 2005}, {\em D14},
2263--2270.
%
%
\bibitem{Padm4}
Padmanabhan,T. The Holography of gravity encoded in a relation
between entropy, horizon area and action for gravity.
 \emph{Gen.Rel.Grav} {\bf 2002}, {\em 34}, 2029--2035.
%
%
\bibitem{Padm5}
Padmanabhan,T. Holographic Gravity and the Surface term in the
Einstein-Hilbert Action. \emph{Braz.J.Phys} {\bf 2005}, {\em 35},
362--372.
%
%
\bibitem{Padm6}
Padmanabhan,T. Gravity: A New holographic perspective.
 \emph{Int.J.Mod.Phys.D} {\bf 2006}, {\em 15}, 1659--1676.
%
%
\bibitem{Padm7}
Mukhopadhyay,A.; Padmanabhan,T. Holography of gravitational action
functionals. \emph{Phys.Rev.D} {\bf 2006}, {\em 74}, 124023.
%
%
\bibitem{Padm8}
Padmanabhan,T. Dark energy and gravity. \emph{Gen.Rel.Grav} {\bf
2008}, {\em 40}, 529--564.
%
%
\bibitem{Padm9}
Padmanabhan,T.; Paranjape,A. Entropy of null surfaces and dynamics
of spacetime. \emph{Phys.Rev. D}  {\bf 2007}, {\em 75}, 064004.
%
%
\bibitem{Padm10}
Padmanabhan,T. Gravity as an emergent phenomenon: A conceptual
description. \emph{International Workshop and at on Theoretical
High Energy Physics (IWTHEP 2007)}, AIP Conference Proceedings
939, American Institute of Physics, New York, 2007; pp. 114--123.
%
%
\bibitem{Padm11}
Padmanabhan,T. Gravity and the thermodynamics of horizons.
 \emph{Phys.Rept}  {\bf 2005}, {\em 406}, 49--125.
%
%
\bibitem{Padm12}
 Paranjape,A.; Sarkar, S.; Padmanabhan,T.
Thermodynamic route to field equations in Lancos-Lovelock gravity.
 \emph{Phys.Rev. D} {\bf 2006}, {\em 74}, 104015.
%
%
\bibitem{Jac1}
Jacobson,T. Thermodynamics of space-time: The Einstein equation of
state.  \emph{Phys. Rev. Lett} {\bf 1995}, {\em 75}, 1260--1263.
%
%
\bibitem{Bek1}
Bekenstein,J.D. Black Holes and Entropy.  \emph{Phys.Rev.D}
{1973}, {\em 7}, 2333--2345.
%
%
\bibitem{Hawk3}
Hawking,S. Black Holes and Thermodynamics. {\emph Phys.Rev. D}
{\bf 1976},{\em 13}, 191--204.
%
%
\bibitem{acs}
 Adler,R. J.; Chen,P.; Santiago, D. I.
 The generalized uncertainty principle and black hole
 remnants. \emph{Gen.Rel.Grav.} {\bf
2001}, {\em 13}, 2101-2108.
%
%
\bibitem{ch}
Custodio, P. S.; Horvath, J. E. The Generalized uncertainty
principle, entropy bounds and  black hole (non)evaporation in a
thermal bath. \emph{Class.Quant.Grav.} {\bf 2003}, {\em 20},
L197-L203.
%
%
\bibitem{Cava:04}
Cavaglia, M.; Das,S. How classical are TeV scale black holes?
 \emph{Class.Quant.Grav.} {\bf 2004}, {\em 21}, 4511--4523.
%
%
\bibitem{Bole:05}
Bolen, B.; Cavaglia,M. (Anti-)de Sitter black hole thermodynamics
and the generalized uncertainty principle.
 \emph{Gen.Rel.Grav.} {\bf 2005}, {\em 37}, 1255--1263.
%
%
\bibitem{mv}
Medved, A.J.M.; Vagenas, E.C. When conceptual worlds collide: The
GUP and the BH entropy. \emph{Phys. Rev. D} {\bf 2004}, {\em 70},
124021.
%
%
\bibitem{Park}
 Park, M.-I. The Generalized Uncertainty Principle in (A)dS Space and the
 Modification of Hawking Temperature from the Minimal Length. \emph{Phys.Lett.B} {\bf 2008},
 {\em 659}, 698--702.
%
%
\bibitem{Kim1}
 Kim,Wontae.; Son,Edwin J.; Yoon, Myungseok.
Thermodynamics of a black hole based on a generalized uncertainty
principle. \emph{JHEP} {\bf 2008}, {\em 08}, 035.
%
%
\bibitem{Nou}
~Nouicer,K. Quantum-corrected black hole thermodynamics to all
orders in the Planck length. \emph{Phys.Lett B} {\bf 2007}, {\em
646}, 63--71.
%
%
\bibitem{Padm13}
Padmanabhan,T. Thermodynamical Aspects of Gravity:
 New insights.Rep. Prog. Phys. 73 (2010) 046901, arXiv:0911.5004.
%
%
\bibitem{Patil}
 Patil,S. P. Degravitation, Inflation and the Cosmological Constant as an
 Afterglow. \emph{JCAP} {\bf 2009}, {\em 0901}, 017.
%
%
\bibitem{Park1}
Park, Mu-in. The Black Hole and Cosmological Solutions in IR
modified Horava Gravity. \emph{JHEP} {\bf 2009}, {\em 0909}, 123.
%
%
\bibitem{Rub}
Rubakov, V. A.; Tinyakov,P. G. Infrared-modified gravities and
massive gravitons. \emph{Phys.Usp} {\bf 2008}, {\em 51}, 123.,
759-792; Nikiforova,V.; Randjbar-Daemi, S.; Rubakov V. Infrared
Modified Gravity with Dynamical Torsion. \emph{Phys.Rev.D} {\bf
2009}, {\em 80}, 124050.
%
%
\bibitem{Balesku1}
Balesku, R. Equilibruim and Nonequilibruim Statistical
Mechanics,v.1,A Wiley Interscience Publications, New
York-London-Sydney-Toronto, 1975.
%
%
\bibitem{Bazarov}
Bazarov, I.P. Thermodynamics,Moskow, Press "Higher School", 1991.
%
%
\bibitem{Gyarm}
Gyarmati, I. Non-Equilibruim Thermodynamics. Field Theory and
Varitional Principles, Springer-Verlag, Berlin-Heidelberg-New
York, 1974.
%
%
\bibitem{Hooft1}
Hooft, G. 'T. Dimensional reduction in quantum gravity.Essay
dedicated to Abdus Salam \emph{gr-qc/9310026}, 15pp.
%
%
\bibitem{Hooft2}
Hooft, G. 'T. The Holographic Principle,
\emph{hep-th/0003004},15pp.; L.Susskind, The World as a hologram.
\emph{J. Math. Phys} {\bf 1995}, {\em 36}, 6377--6396.
%
%
\bibitem{Bou1}
Bousso, R. The Holographic principle. \emph{Rev. Mod. Phys} {\bf
2002}, {\em 74}, 825--874.
%
%
\bibitem{Bou3}
Bousso, R. A Covariant entropy conjecture. \emph{JHEP} {\bf 1999},
 {\em 07}, 004.
%
%
\bibitem{Bal}
Balazs,C.; Szapudi,I. Naturalness of the vacuum energy in
holographic theories. \emph{hep-th/0603133}, 4pp.
%
%
\bibitem{Sussk1}
Fischler,W.; Susskind,L. Holography and cosmology.
\emph{hep-th/9806039},7pp.
%
%
\bibitem{Cai1}
Cai,R.-G.; Kim,S.P. First law of thermodynamics and Friedmann
equations of Friedmann-Robertson-Walker universe.
 \emph{JHEP } {\bf 2005}, {\em 02}, 050.
%
%
\bibitem{Min1}
Jejjala, V.; Kavic, M.; Minic, D. Time and M-theory.
 \emph{Int. J. Mod. Phys. A}
 {\bf 2007}, {\em 22}, 3317--3405.
%
%
\bibitem{Min2}
Jejjala, V.; Kavic, M.; Minic, D. Fine structure of dark energy
and new physics. \emph{Adv. High Energy Phys.} {\bf 2007}, {\em
2007}, 21586.
%
%
\bibitem{Min3}
Jejjala, V.; Minic, D. Why there is something so close to nothing:
Towards a fundamental theory of the cosmological constant.
\emph{Int.J.Mod.Phys.A} {\bf 2007},
 {\em 22}, 1797-1818.
%
%
\bibitem{Min4}
Jejjala, V.; Minic, D.; Tze, C-H. Toward a background independent
quantum theory of gravity. \emph{Int. J. Mod. Phys. D}, {\bf
2004}, {\em 13}, 2307--2314.
%
%
\bibitem{Kolb}
 Kolb E., Turner M. The Early Universe , Reading, Addison
Wesley, 1990; Baumann D. TASI Lectures on Inflation
//arXiv:0907.5424. P. 1---160.
%
%
\bibitem{Polyak}
Polyakov A. M. Decay of Vacuum Energy. \emph{Nucl.Phys.B.}, {\bf
2010}, {\em 834},  316--329.
%
%
\end{thebibliography}
\end{document}